\documentclass[journal=jacsat,manuscript=article]{achemso}

\usepackage[version=3]{mhchem} 
\usepackage{graphicx}
\usepackage{dcolumn}
\usepackage{bm}
\usepackage{braket}
\usepackage{enumitem}
\usepackage{color}
\usepackage{booktabs} 
\usepackage{newfloat}
\usepackage{algcompatible}
\usepackage[size=small]{caption}
\usepackage{etoolbox}
\usepackage{hyperref}

\author{Denis G. Artiukhin}
\affiliation[Freie Universität Berlin]
{Institut für Chemie und Biochemie, Freie Universität Berlin, \\ Arnimallee 22, 14195 Berlin, Germany.}
\email{denis.artiukhin@fu-berlin.de}

\title[Initial Guesses]
  {Initial Guesses for Multicomponent Mean-Field Methods: Assessment and New Developments}

\abbreviations{}
\keywords{}

\begin{document}

\begin{abstract}
	
The convergence of self-consistent field equations in mean-field nuclear--electronic orbital methods strongly depends on the choice of initial guesses for quantum nuclei. Although several such guesses have been proposed in the literature, a systematic comparison of their performance as well as attempts of constructing novel approximations based on model tasks of quantum mechanics were not reported to date. In this work, we address both issues by introducing novel nuclear initial guesses derived from the analytical solutions of the three-dimensional quantum harmonic oscillator and benchmarking them against existing approaches. We demonstrate that the isotropic variant of our guess outperforms existing approximations in nuclear--electronic orbital density functional theory calculations employing the simultaneous self-consistent field convergence protocol. Although our guess requires the computation of partial Hessians, we demonstrate that these can be evaluated with low-cost methods without affecting the accuracy of resulting protonic density matrices. Our results demonstrate that the proposed guess is robust and efficient and could provide a route to improved convergence in mean-field nuclear--electronic orbital computations.

\end{abstract}

\section{Introduction} \label{sec:intro}

The Roothaan--Hall equations~\cite{Roothaan1951,Hall1951} and their variants are among the most frequently solved problems in electronic-structure computational chemistry. They constitute the foundation of all Kohn--Sham Density Functional Theory (DFT) computations and, being employed within Hartree--Fock (HF) theory, define the reference state employed in the majority of correlated wave-function-based methods. In the atomic orbital (AO) representation, these equations take the form of a generalized eigenvalue problem,
\begin{equation} \label{eq:roothaan-hall}
	\mathbf{FC} = \mathbf{SC\varepsilon}
\end{equation}
where $\mathbf{F}$ is the Fock matrix, $\mathbf{C}$ is the matrix of molecular orbital (MO) coefficients, $\mathbf{S}$ is the AO overlap matrix, and $\mathbf{\varepsilon}$ is the diagonal matrix of orbital energies. Because the Fock matrix $\mathbf{F}$ dependents on the density matrix $\mathbf{P}$ and, therefore, implicitly on the unknown MO coefficients $\mathbf{C}$, Eq.~(\ref{eq:roothaan-hall}) has to be solved iteratively. The resulting self-consistent procedure requires an initial guess, which can be specified in terms of the Fock matrix $\mathbf{F}$, the density matrix $\mathbf{P}$, or the MO coefficient matrix 
$\mathbf{C}$. Since the original formulation of the Roothaan--Hall equations, a large variety of initial guesses has been proposed, which are nowadays routinely employed in electronic-structure computations. Among those, to name but a few, are the one-electron or core guess, extended H\"uckel method~\cite{Hoffmann1963}, superposition of atomic densities (SAD) approach~\cite{amloef1982,lenthe2006}, and the superposition of atomic potentials (SAP) method~\cite{Lehtola2019}.

In the mean-field Nuclear--Electronic Orbital (NEO) approaches, such as NEO-HF~\cite{webb2002} and NEO-DFT~\cite{pak2007,chakraborty2008,chakraborty2009,sirjoosingh2011,sirjoosingh2012}, a few selected nuclei, typically protons, are treated quantum mechanically and on the very same footing with electrons, giving rise to two sets of self-consistent field (SCF) equations,
\begin{align}
	\mathbf{F}^{\mathrm{e}} \mathbf{C}^{\mathrm{e}} &= \mathbf{S}^{\mathrm{e}} \mathbf{C}^{\mathrm{e}} \mathbf{\varepsilon}^{\mathrm{e}}, \label{eq:neo_el_scf} \\ 
	\mathbf{F}^{\mathrm{p}} \mathbf{C}^{\mathrm{p}} &= \mathbf{S}^{\mathrm{p}} \mathbf{C}^{\mathrm{p}}\mathbf{\varepsilon}^{\mathrm{p}} \label{eq:neo_prot_scf},
\end{align}
where the former expression is defined for the system of electrons and the latter is given for quantum protons as indicated by superscripts ``e'' and ``p'', respectively. 
Here, each of the Fock matrices $\mathbf{F}^{\mathrm{e}}$ and $\mathbf{F}^{\mathrm{p}}$ depends on both the electronic and protonic densities $\mathbf{P}^{\mathrm{e}}$ and $\mathbf{P}^{\mathrm{p}}$, coupling the two set of equations.
While the aforementioned initial guesses remain applicable for the system of electrons, a distinct class of protonic guesses has to be introduced for the quantum-mechanically treated protons. To date, this topic received very limited attention in the literature and, to the best of our knowledge, only three types of such initial guesses were reported.

In close analogy to the one-electron initial guess, Mata and co-workers proposed a core guess for quantum protons in Ref.~\citenum{Hasecke2023}. In this approach, the elements of the Fock matrix $\mathbf{F}^{\mathrm{p}}$ are approximated as,
\begin{equation}
	F^{\mathrm{p}}_{\mu \nu} \approx h^{\mathrm{P}}_{\mu \nu} - \sum_{\alpha \beta}^{N_{\mathrm{bf}}^{\mathrm{e}}} P_{\alpha \beta}^{\mathrm{e}} J^{\mathrm{pe}}_{\alpha \beta},
\end{equation}
where the first term $h^{\mathrm{P}}_{\mu \nu}$ on the right-hand side is the one-proton contribution accounting for the kinetic energy of quantum protons and quantum proton--classical nuclei interaction, and the second term being the Coulomb interaction of quantum protons and electrons. The latter depends on the electronic density matrix $\mathbf{P}^{\mathrm{e}}$ and the corresponding matrix of Coulomb interaction intergrals, with indices $\alpha$ and $\beta$ running over the number of electronic basis functions $N_{\mathrm{bf}}^{\mathrm{e}}$. It was demonstrated that the use of this approximation can result in several quantum nuclei being localized at a single center. To prevent this from happening, a block-diagonalization procedure was proposed, where parts of the Fock matrix corresponding to each quantum protons are diagonalized separately~\cite{Hasecke2023}. The combination of this procedure and Coulomb coupling elements being considered made this approach applicable to a broad range of molecules.

Since protonic densities are much more localized in space compared to those of electrons and are often of nearly spherical shape, they can be approximated by the density of an s-type orbital. The first protonic guess of this type was reported by Li and co-workers in Ref.~\citenum{Aodong2022}, where it was proposed to occupy the tightest basis function of a given protonic basis set for each quantum proton. As an alternative, Lehtola and co-workers proposed to employ a single 1s-type basis function with a fitted value of the exponent (equal to 16$\sqrt{2}$~bohr$^{-2}$) in Ref.~\citenum{Lehtola2025}. While the former approach depends on the choice of the protonic basis set, the latter requires re-fitting of the exponent value for computations of heavier quantum nuclei.

The third type of a protonic initial guess was proposed by Reiher and co-workers in Ref.~\citenum{Feldmann2023} in close analogy to the electronic SAD approximation. In this so-called superposition of nuclear densities (SND) guess, the total nuclear density matrix $\mathbf{P}^\mathrm{p}$ is constructed as a direct sum of density matrices of individual nuclei, which are obtained in separate atomic SCF computations. The latter employ a Hamiltonian with subtracted center-of-mass contributions~\cite{Muolo2020} making the SND guess somewhat more elaborate than those described before. SND was reported to perform better than the core guess, however, the latter variant lacked the proton--electron Coulomb interaction contributions and was not implemented in conjunction with the block-diagonalization procedure.

Although the aforementioned protonic guesses have been tested on various molecular systems and shown to produce satisfactory SCF convergence behavior, to the best of our knowledge, their performance has not been systematically compared. Furthermore, since the protonic density obtained within NEO approaches can be considered as a vibrational density of the quantum-mechanically treated protons, it is somewhat surprising that protonic initial guesses derived from analytic solutions of simple quantum-mechanical model problems, such as the harmonic oscillator, have not been explored in this context. 
The goal of this paper is therefore to address both of these issues by introducing new protonic initial guesses and comparing their performance with existing approaches.

This work is organized as follows. The theoretical background and the construction of two initial guesses based on analytical solutions of the three-dimensional quantum harmonic oscillator (HO) problem are introduced first. Subsequently, the computational protocol is described, followed by a comparison of the new and existing protonic initial guesses. The work concludes with an outline of main findings.

\section{Theory} \label{sec:theory}
A model system XH is considered, where a hydrogen atom H (protium, deuterium, or tritium) moves with respect to the rest of the molecule X. We intend to describe the hydrogen nucleus quantum mechanically with the NEO-HF~\cite{webb2002} or NEO-DFT~\cite{pak2007,chakraborty2008,chakraborty2009,sirjoosingh2011,sirjoosingh2012} approaches. In this case, the generation of initial SCF guesses can be performed by
\begin{enumerate}
	\item Evaluating and diagonalizing a partial Hessian matrix for the hydrogen nucleus;
	\item Solving the 3D quantum HO problem;
	\item Projecting the analytical HO ground-state wave function onto the chosen nuclear basis set.
\end{enumerate}

In the following, the theory behind the 3D-anisotropic HO is briefly outlined in Section \nameref{sec:HO}. In Section \nameref{sec:projection}, we present the wave function projection technique and generalize our approach towards multiple quantum nuclei. 
Finally, in Section \nameref{sec:overlaps} we discuss a technical issue of computing overlap elements between anisotropic and isotropic Gaussian functions, which are required in the projection step, as well as present further simplifications of our approach and its relation to the existing 1s-type initial guess.

\subsection{3D-Anisotropic Harmonic Oscillator} \label{sec:HO} 
Let us assume that the coordinates of the quantum hydrogen nucleus are $\mathbf{R}=(X,Y,Z)$ with the equilibrium position under the Born--Oppenheimer approximation being $\mathbf{R}_\mathrm{e} =(X_\mathrm{e},Y_\mathrm{e},Z_\mathrm{e})$. The frame can be centered to the equilibrium position of the hydrogen $\mathbf{Q} = (\mathbf{R}-\mathbf{R}_e)$. In the given frame, we can compute the partial 3$\times$3 Hessian $\mathbf{H}$ for the motion of this hydrogen within the molecule, composed of elements $H_{\alpha \beta} = \partial_{\alpha \beta}^2V|_\mathrm{e}$, where $\alpha,\beta \in\{ X,Y,Z\}$, index ``e'' denotes that the second partial derivative is taken at the equilibrium geometry at the given electronic theory approximation, and $V$ is the potential energy surface computed at some low-cost electronic approximation.   

The Hessian $\mathbf{H}$ can be diagonalized using an orthogonal transformation $\mathbf{L}$ to the new displacement coordinates, 
\begin{equation}
\label{eq:trans_coords}
\mathbf{r}=(x,y,z)=\mathbf{L}\mathbf{Q}
\end{equation}
to the form $\mathbf{H}_\mathrm{d}=\mathbf{L}\mathbf{H}\mathbf{L}^\mathrm{T}=\mu\cdot \operatorname{diag}(\omega_x^2, \omega_y^2, \omega_z^2)$, where $\mu$ is the mass of hydrogen, and $\omega_\xi$, $\xi\in\{x,y,z\}$, are the angular frequencies (related to the frequencies $\nu$ and expressed in inverse units of length $\tilde{\nu}$ as $\omega = 2\pi \nu = 2\pi c \tilde{\nu}$, where $c$ is the speed of light) along given nuclear displacements from the equilibrium position.

In this way, we express the harmonic potential energy of the quantum hydrogen nucleus as~\cite{pauling},
\begin{equation} \label{eq:ho_potential}
V(\mathbf{r})=\frac{1}{2} \mathbf{r}^\mathrm{T}\mathbf{H}_\mathrm{d}\mathbf{r}  = \sum_{\xi\in\{x,y,z\}} \frac{\mu \omega_{\xi}^2 \xi^2}{2} \ .
\end{equation}
The Schr\"odinger equation of a particle moving in the potential given in Eq.~(\ref{eq:ho_potential}) can be solved analytically, resulting in the wave function of the form
\begin{align}
\psi(\mathbf{r})=\psi_x(x)\psi_y(y)\psi_z(z),
\end{align}
where each function $\psi_{\xi}(\xi)$ is the solution of the one-dimensional HO. We are interested in the ground vibrational state, as this is the target state for the NEO-SCF procedure. Therefore, the wave function $\psi_{\xi}(\xi)$ is given as
\begin{equation}
\psi_{\xi}(\xi)=\left(\frac{2\mu\omega_{\xi}}{h}\right)^{1/4} \exp\left( - \frac{\mu \omega_{\xi} \xi^2}{2\hbar} \right) .
\end{equation}
Here, $h$ is the Planck constant and $\hbar = h/(2\pi)$ is the reduced Planck constant. The final expression for the ground state wave function is an anisotropic Gaussian function of the form,
\begin{equation}\label{eq:hofinal}
\psi(\mathbf{r})=\left(\frac{2\mu}{h}\right)^{3/4}(\omega_x\omega_y\omega_z)^{1/4}  \exp\left(- \frac{\mu}{2\hbar}\sum_{\xi\in\{x,y,z\}} \omega_\xi \xi^2\right). 
\end{equation}
Following derivations from Ref.~\citenum{Tikhonov2023}, we rewrite the expression in the exponent as
\begin{equation}
\label{eq:exponent_expr}
-\frac{\mu}{2\hbar}\sum_{\xi\in\{x,y,z\}} \omega_\xi \xi^2 = -\frac{1}{4} \mathbf{r}^\mathrm{T}\bm{\Sigma}^{-1}\mathbf{r} \ , 
\end{equation}
where $\bm{\Sigma}^{-1}$ is the diagonal inverse variance matrix with elements $(\bm{\Sigma}^{-1})_{\xi\xi} = {2\mu}\omega_\xi/{\hbar}$, and express our wave function from Eq.~(\ref{eq:hofinal}) in terms of the original displacements $\mathbf{Q}$,
\begin{equation}\label{eq:hofinal_Q}
\psi(\mathbf{Q})=\left(\frac{2\mu}{h}\right)^{3/4}(\omega_x\omega_y\omega_z)^{1/4}   \exp\left(- \frac{1}{4}\mathbf{Q}^\mathrm{T}\mathbf{L}^\mathrm{T}\bm{\Sigma}^{-1}\mathbf{L}\mathbf{Q} \right).
\end{equation}
In the following, this expression is employed to approximate the ground state nuclear wave function for the quantum hydrogen nucleus.

It should be noted that the aforementioned procedure of diagonalizing only a sub-block of the Cartesian Hessian can be referred to as the Partial Hessian Vibrational Analysis (PHVA) and was successfully applied in the literature for computations of vibrational enthalpy and entropy changes in chemical reactions~\cite{phva2002}. In the context of the current work, it is a rather crude approximation, which describes the movement of the hydrogen atom H relative to the infinitely heavy moiety X and results in angular frequencies $\omega_\xi$ being contaminated by rotational and translational degrees of freedom. The consequences of using this approximation for generating an initial guess are further discussed in Section \nameref{sec:comput_details}.

\subsection{Basis Set Projections} \label{sec:projection}
As seen from Eq.~(\ref{eq:hofinal_Q}), the quantum hydrogen nucleus is described with a single anisotropic Gaussian, whereas isotropic Gaussian functions are used for constructing AOs in the NEO-HF and NEO-DFT approaches. In order to make our method practical, a basis set projection technique can be applied~\cite{fink1973,soriano2014}. 
This technique is routinely used in several program packages, for example, in \textsc{Serenity}~\cite{unsleber2018,niemeyer2023,serenity_zenodo} and \textsc{PySCF}~\cite{pyscf1,pyscf2}, often in conjunction with the electronic SAD initial guess~\cite{amloef1982,lenthe2006}. In this approach, a projection operator onto the user specified AO basis set is constructed. Applying this operator to an HO function leads to its representation as a linear combination of protonic AO basis functions. In this work, we use this technique to directly construct the density matrix $\mathbf{P}_{\mathrm{2}}$ defined in the user specified protonic set (basis 2) via the given density matrix $\mathbf{P}_{\mathrm{1}}$ corresponding to the HO function (basis 1),
\begin{equation}
\mathbf{P}_{2}  = \mathbf{U} \mathbf{P}_{1} \mathbf{U}^{\mathrm{T}},
\end{equation} 
where the transformation matrix $\mathbf{U}$ is defined by
\begin{equation} \label{eq:transformation_matrix}
\mathbf{U} = \mathbf{S}_{22}^{-1} \mathbf{S}_{21},
\end{equation}
where $\mathbf{S}_{22}$ is the AO overlap matrix and  
$\mathbf{S}_{21}$ contains overlaps of our HO function with AO basis functions. Detailed derivations of these expressions are given in Section S1.1 of the Supporting Information (SI).
Therefore, the problem reduces to computing the overlap matrices $\mathbf{S}_{22}$ and $\mathbf{S}_{21}$, inverting $\mathbf{S}_{22}$, and constructing $\mathbf{U}$. The resulting density matrix $\mathbf{P}_2$ is then employed as the initial guess in nuclear SCF. In case of multiple quantum hydrogen nuclei being present in the molecular system, HO wave functions need to be constructed and projected independently for each quantum nucleus. Then, the block diagonal density matrix of the entire quantum nuclear system $\widetilde{\mathbf{P}}_2$ is constructed from matrices $\mathbf{P}_2$ of single nuclei.

\subsection{Computing Overlaps} \label{sec:overlaps}
Computations of AO overlaps $\mathbf{S}_{22}$ are implemented and routinely carried out in many quantum chemical program packages or, alternatively, performed using freely-available external libraries such as \textsc{LibInt}~\cite{Libint2}. However, this is not the case for overlap matrix elements $\mathbf{S}_{21}$ between anisotropic- and isotropic-type Gaussian functions required in Eq.~(\ref{eq:transformation_matrix}). Therefore, we derived and implemented the corresponding analytical expressions.

Denoting the elements of the vector $\mathbf{Q}$ as $Q_x$, $Q_y$, and $Q_z$, we can write the expression for a Cartesian Gaussian-type AO as~\cite{Obara86},
\begin{equation} \label{eq:atomic_orbital_cart}
\chi_{lmn}(\mathbf{Q})=N_{lmn} Q_x^l Q_y^m Q_z^n \exp(-\zeta \mathbf{Q}^{\mathrm{T}}  \mathbf{Q} ),
\end{equation}
where the normalization constant $N_{lmn}$ is given as
\begin{equation}
N_{lmn} = \left( \frac{2 \zeta}{\pi}  \right)^{3/4} \frac{(4 \zeta)^{(l+m+n)/2}}{ \sqrt{ (2l-1)!! (2m-1)!! (2n-1)!!  } } 
\end{equation}
and depends on the three components $l$, $m$, and $n$ of the angular momentum, and $\zeta$ is a positive real number.
The overlap integral $\braket{\psi | \chi}$ can then be expressed as
\begin{equation} \label{eq:overlap_initial}
\braket{\psi | \chi_{lmn}} = N_{\psi} N_{lmn}     \int_{-\infty}^{+\infty}  Q_x^l Q_y^m Q_z^n \exp\left(-\frac{1}{2}\mathbf{Q}^{\mathrm{T}} \mathbf{A}^{-1} \mathbf{Q} \right) \mathrm{d}\mathbf{Q},  
\end{equation}
where $N_{\psi} = (2 \mu/h)^{3/4} (\omega_x \omega_y \omega_z)^{1/4}$ is the normalization constant from Eq.~(\ref{eq:hofinal_Q}), $\mathbf{A}^{-1} = (\frac{1}{2} \mathbf{L}^\mathrm{T}\bm{\Sigma}^{-1}\mathbf{L} + 2\mathbf{I} \zeta)$, and $\mathbf{I}$ is the identity matrix. This integral can be solved analytically. Due to the symmetric integration limits in Eq.~(\ref{eq:overlap_initial}) and the exponential expression being an even function, only even values of $l+m+n$ lead to non-zero contributions, i.e., only s-, d-, and g-type AOs need to be considered. The final expression reads,
\begin{equation}
\braket{\psi | \chi_{lmn}} = \frac{(2 \pi)^{3/2}}{\sqrt{\mathrm{det(\mathbf{A}^{-1})}}}  N_{\psi} N_{lmn} M_{lmn},
\end{equation}
where $M_{lmn}$ is equal to one for s-type functions and is expressed via elements of matrix $\textbf{A}$ for higher angular momenta. Analytical expressions for values $M_{lmn}$ are derived and tabulated in Section S1.2 in the SI.

With the overlap matrix elements being derived, our HO anisotropic initial guess (referred to as ``HOa'' in the following) is fully defined and can be used in practical computations. From a theoretical standpoint, it could potentially provide a faster SCF convergence in cases of strongly anisotropic proton densities. Furthermore, since no empirical parameters are employed and the particle mass $\mu$ is specified on the input, HOa can be applied for heavier hydrogen isotopes or even different nuclei, such as helium or lithium, without any theory or code modifications. It should be noted, however, that working expressions presented are explicitly derived for the case of Cartesian basis functions and an additional transformation is required to use it in conjunction with spherical AOs~\cite{Schlegel1995}. Such a transformation is not considered in this work and HOa is tested exclusively with Cartesian basis functions.

A further simplification of this approach and a direct connection with the existing isotropic 1s-type initial guess for protons~\cite{Aodong2022,Lehtola2025} can trivially be established by employing a single $\omega$ value, which can be chosen, for example, as the arithmetic mean $(\omega_x + \omega_y + \omega_z)/3$ or as $\max(\omega_x, \omega_y, \omega_z)$. In this case, the inverse variance matrix $\bm\Sigma^{-1}$ becomes $(2 \mu \omega/ \hbar) \mathbf{I}$ and due to the fact that $\mathbf{L}^{\mathrm{T}} \mathbf{L} = \mathbf{I}$,
the corresponding expression for the wave function $\psi(\mathbf{Q})$ reads
\begin{equation} \label{eq:isotropic_wavefunc}
\psi (\mathbf{Q}) = \left( \frac{2 \mu \omega}{h}\right)^{3/4} \exp \left( - \frac{\mu \omega}{ 2 \hbar}   \mathbf{Q}^{\mathrm{T}} \mathbf{Q}  \right). 
\end{equation} 
Comparing Eqs.~(\ref{eq:atomic_orbital_cart}) and (\ref{eq:isotropic_wavefunc}), one can see that the latter is a Gaussian 1s-type AO function.
Therefore, the task of computing overlap integrals $\braket{\psi | \chi_{lmn}}$ reduces to evaluations of integrals of the form $\braket{ \chi_{000} | \chi_{lmn}}$ and can be carried out by employing standard libraries such as \textsc{LibInt}. Furthermore, this simplified version is no longer restricted to be used with Cartesian AOs only. Contrary to the existing 1s-type initial guess for protons~\cite{Aodong2022,Lehtola2025}, in this approach the value of the exponent $\zeta = \mu \omega / 2 \hbar$ is not found empirically by means of fitting but directly computed from the angular frequency $\omega$ and the particle mass $\mu$ providing a higher flexibility and applicability to a larger range of molecules. In the following, we denote this isotropic HO initial guess as ``HOi''.

\section{Computational Details} \label{sec:comput_details}

To test various initial guesses, the NEO-HF~\cite{webb2002} and NEO-DFT~\cite{pak2007,chakraborty2008,chakraborty2009,sirjoosingh2011,sirjoosingh2012} methods were implemented in the developer version of the \textsc{Serenity}~\cite{unsleber2018,niemeyer2023,serenity_zenodo} program package. The correctness of the program implementation was assessed by comparing results with those generated with the Q-Chem software~\cite{QChem}. Details of this analysis are presented in Section~S2 of the SI.

The test set of molecules employed in this work is shown in Tab.~\ref{tabl:test_set}, where quantum mechanically treated protons are given in bold fonts. A single protons was treated as quantum in the first fourteen molecules from the set, whereas multiple protons were quantum in the last six molecules. Molecular structures were optimized with \textsc{Serenity} employing the Hartree--Fock (HF) and Density Functional Theory (DFT) methods. In both cases, the augmented correlation--consistent polarized valence triple-$\zeta$ aug-cc-pVTZ
basis set~\cite{Dunning1989} was used. Integral prescrining threshold was set to 1.0$\times 10^{-14}$~a.u. The SCF procedure was initiated with the superposition of atomic densities (SAD) initial guess~\cite{amloef1982,lenthe2006} and converged tightly with the thresholds for the energy, density matrix elements, and the direct inversion in the iterative subspace (DIIS) error being set to 
1.0$\times 10^{-8}$,
1.0$\times 10^{-8}$,
and 1.0$\times 10^{-7}$ a.u., respectively. In DFT calculations, the exchange--correlation functional PBE~\cite{Perdew1996} was employed in conjunction with Becke integration grids~\cite{Becke1988,Treutler1995} of quality 5 for SCF iterations and 6 for the final energy evaluation. Density fitting procedures were not employed. This set of settings was consistently applied in all structure optimization runs, computations of Hessian matrix blocks, and in subsequent NEO-HF and NEO-DFT calculations unless stated otherwise. The optimized molecular structures were confirmed as minima using a normal mode
analysis. 

\begin{table}[h!]
	\small
	\center
	\caption{Test set of molecules employed in this work. Hydrogen nuclei, which are treated quantum mechanically, are given in bold fonts.}
	\label{tabl:test_set}
	\begin{tabular}{c}
		\hline
		\hline
		F\textbf{H}F$^-$, \,   H$_2$O$\cdots$\textbf{H}F, \,  
		H$_2$O$\cdots$\textbf{H}Cl, \,    H$_2$O$\cdots$\textbf{H}OH, \, CH$_3$OH$\cdots$\textbf{H}OH, \,     
		H\textbf{H}NO, \,  \\ HO\textbf{H}, \, HS\textbf{H}, \,  \textbf{H}CN, \, \textbf{H}FCO, \, \textbf{H}NO, \, \textbf{H}OCl, \, \textbf{H}OF, \, \textbf{H}ONO, C\textbf{H$_2$}O, \, \\
		 C\textbf{H$_3$}F, \, C$_2$\textbf{H$_4$}, \,
		\textbf{H$_2$}O$\cdots$\textbf{H}F, \,  
		\textbf{H$_2$}O$\cdots$\textbf{H}Cl, \,    
		  \textbf{H$_2$}O$\cdots$\textbf{H$_2$}O, \,   \\
		\hline
		\hline          
	\end{tabular}                
\end{table}

The HF- and DFT-optimized molecular structures as well as partial Hessians were used in NEO-HF and NEO-DFT computations, respectively. Four initial protonic guesses were implemented and tested: 1s-type guess~\cite{Aodong2022,Lehtola2025} with the value of the exponent being equal to 16$\sqrt{2}$~bohr$^{-2}$, a variant of the core guess by Mata and co-workers~\cite{Hasecke2023}, as well as HOa and HOi guesses discussed in this work and given in Eqs.~(\ref{eq:hofinal_Q}) and (\ref{eq:isotropic_wavefunc}), respectively. As mentioned before, the implementation of the SND guess is associated with additional technical complications due to the need to remove center-of-mass contributions. Therefore, it is not considered in this work.
Basis sets from the PB family~\cite{Yu2020} by Hammes--Schiffer and co-workers were applied. We note, however, that these sets were recently shown
not to form a systematically convergent hierarchy. For more discussion on this topic, the interested reader is referred to Ref.~\citenum{nikkanen2026}. To avoid problems of protonic basis sets becoming linear dependent, calculations employing Cartesian AOs were performed with PB4 basis sets only (more specifically, PB4-D, PB4-F1, and PB4-F2), whereas PB4 and PB5 (i.e., PB5-D, PB5-F, and PB5-G) families of basis sets were used in case of spherical AOs. The canonical orthogonalization threshold was set to 1.0$\times 10^{-7}$~a.u.\ for both protonic and electronic tasks. In order to generate an initial guess for the electronic problem within NEO-HF and NEO-DFT, SCF computations of molecules without quantum protons were performed first using the SAD guess~\cite{amloef1982,lenthe2006}. Subsequently, the converged electronic densities were used as starting points for computations of molecules featuring quantum protons. We note that a potentially improved algorithm could be established by performing another electronic SCF step in the presence of coupling terms constructed from a guessed protonic density before the first protonic SCF procedure. This algorithmic refinement will be addressed in future work.
In NEO-DFT computations, the electron--proton correlation functional EPC172~\cite{Brorsen2017} from the LibXC library~\cite{LibXC} was employed.

Mean-field NEO approaches are known to have slower and more challenging SCF convergence compared to their electronic structure counterparts~\cite{Bochevarov2004,Aodong2022,Feldmann2023}. It is, therefore, suggested to use second-order optimization algorithms for these methods~\cite{Feldmann2023,Greiner2026}. However,
since testing different sophisticated SCF optimization procedures goes beyond the scope of this work, a relatively simple approach including the standard Pulay's DIIS procedure~\cite{Pulay1980,Pulay1982} with default for Serenity level-shifting and damping during first SCF iterations was employed in all reported computations. We found this procedure to be sufficient for converging almost all (with only a few exceptions) NEO-SCF computations for molecules containing a single quantum proton. Unfortunately, a considerable part of computations for molecules with multiple quantum protons failed to converge or converged to wrong minima. Therefore, in the following we focus our attention exclusively on molecules with a single quantum proton, whereas present all obtained results in Sections S3--S5 of the SI. We further note that the results obtained for systems with single and multiple quantum protons demonstrate very similar trends, and no clear correlation between the choice of initial guess and SCF convergence issues was observed.

The two NEO-SCF convergence strategies were implemented and used: step-wise and simultaneous~\cite{Aodong2022}.
In the first case, SCF procedures were performed separately for electrons and protons until the change in the total energy drooped below 1.0$\times 10^{-8}$ a.u. These separate SCFs employed same types and values of convergence thresholds as described above for standard HF and DFT. The maximal number of SCF iterations was set to 200, whereas the maximal number of macro-iterations, each composed of one SCF for electrons and one for protons, was set to 50.
In the simultaneous procedure, both electronic and protonic Fock matrices are updated in each DIIS step. In this algorithm, the maximal number of combined SCF iterations was set to 300. As convergence thresholds, the total NEO-SCF energy, the change in protonic and electronic density matrices, and the combined DIIS errors were checked. The thresholds were set to  1.0$\times 10^{-8}$, 1.0$\times 10^{-8}$, and 1.0$\times 10^{-7}$ a.u., respectively. NEO-SCF computations employing the same basis set but different protonic initial guesses were confirmed to converge to same energetic minima with the largest found deviation of about 5$\times 10^{-7}$~a.u. Somewhat larger deviations of up to 2$\times 10^{-5}$~a.u. were found for NEO-DFT, when using different convergence protocols, presumably due to different types of convergence thresholds being employed and additional numerical noise being introduced by integration grids.

Because we employed a rather crude approximation making use of 3$\times$3 partial Hessian matrices and therefore neglecting important mode--mode interactions between the target proton and the rest of the molecule, resulting angular frequencies $\omega_{\xi}$, $\xi = x, y, z$ often featured one overly small value. This led to HO wavefunctions being strongly delocalized along such direction $\xi$ with subsequent SCF convergence issues in calculations employing HOa. To prevent this from happening, values of $\omega_{\xi}$ which are below 1.0$\times 10^{-4}$~a.u.\ were set to the next in magnitude angular frequency. This pragmatic solution ensured numerical stability and kept the anisotropic form of the wave function in the HOa guess. In the case of the HOi approximation, a single angular frequency being equal to $\max( \omega_x, \omega_y, \omega_z)$ was used.

The simplest approach for assessing the quality of initial guesses is to analyze numbers of resulting SCF iterations. This strategy, however, could depend strongly on the type of SCF optimization algorithm used. An alternative approach was proposed in Ref.~\citenum{Lehtola2019} and is based on assessing how close the guessed orbitals are to those of the converged ground state. This criterion reads,
\begin{equation}
	f = \frac{\mathrm{trace}( \mathbf{P}_{\mathrm{guess}} \mathbf{S} \mathbf{P}_{\mathrm{true}} \mathbf{S} )}{N_{\mathrm{p}}},
\end{equation}
where $\mathbf{P}_{\mathrm{guess}}$ and $\mathbf{P}_{\mathrm{true}}$ are guessed and converged protonic density matrices, $\mathbf{S}$ is the AO overlap matrix, and $N_{\mathrm{p}}$ is the number of protons. This measure is defined in the interval from zero to one, where zero describes the worst possible scenario and one corresponds to the best match between the guessed and converged solutions. We note, however, that this measure has a disadvantage of being not fully reliable for non-idempotent density matrices. For further discussion on this topic, we refer the interested reader to Ref.~\citenum{Lehtola2019}. In the current work, both $f$-ranking scores and numbers of SCF iterations are recorded and analyzed.

In Section~\nameref{sec:app_hessians}, smaller basis sets such as the correlation consistent polarized valence double-$\zeta$ cc-pVDZ~\cite{Dunning1989} and minimal STO-3G~\cite{Hehre1969} as well as the less expensive approach GFN2-xTB~\cite{Bannwarth2019} from the \textsc{xtb}~\cite{Bannwarth2020} program were tested for computations of low-cost partial Hessians. After performing a structure optimization and a subsequent computation of a Hessian, only the latter was used in NEO-HF and NEO-DFT computations featuring same computational protocol as described previously, i.e., with larger basis sets and higher-quality molecular structures.

\section{Results and Discussion} \label{sec:results}

\subsection{$f$-Rank Scores} \label{sec:f_scores}

\begin{figure}
	\centering
	\includegraphics[width=0.92\textwidth]{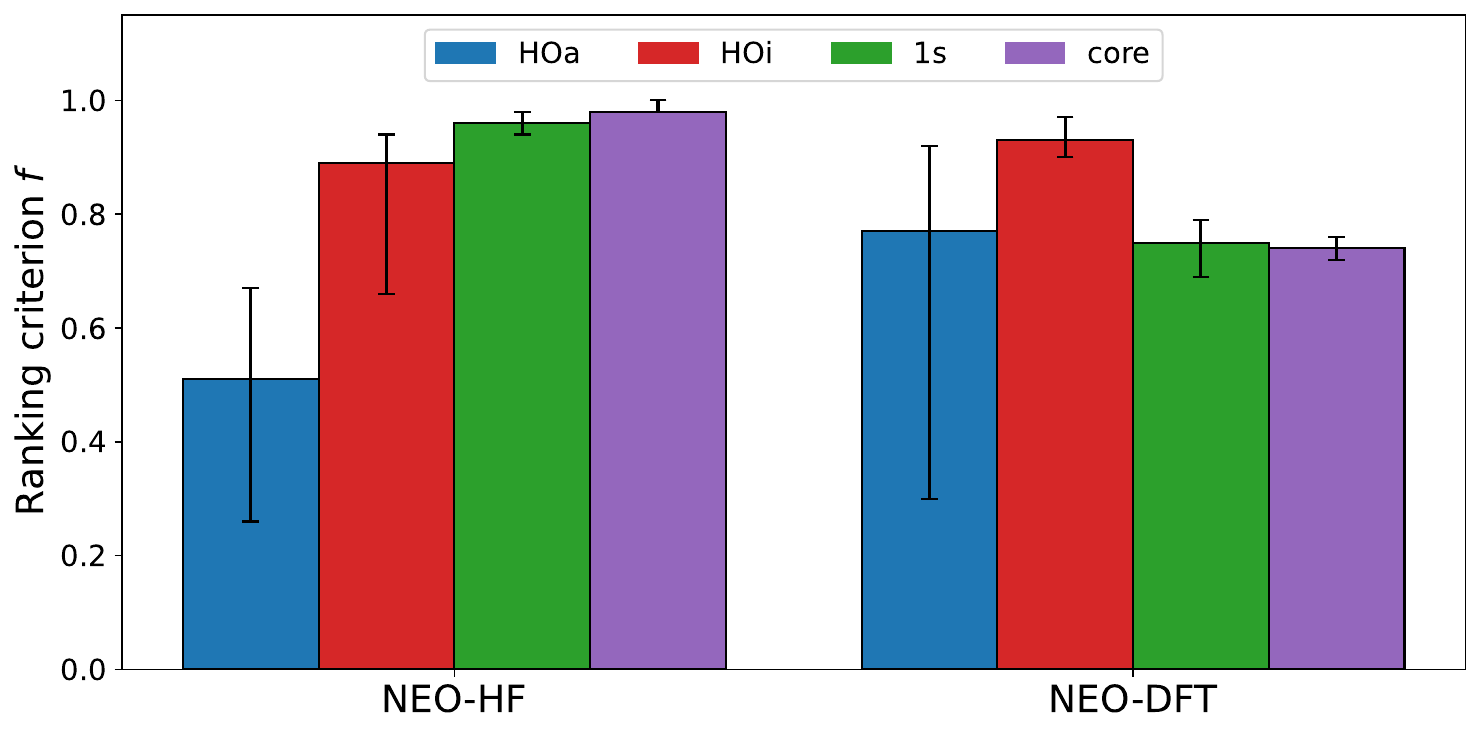}
	\caption{$f$-values computed with NEO-HF and NEO-DFT using the Cartesian PB4-F2 basis set and four different initial guesses for molecules containing a single quantum proton. Bars correspond to $f$-values averaged over all molecules in the set containing a single quantum proton, whereas range bars indicate minimal and maximal $f$-values.}
	\label{fig:f_ranking}
\end{figure}

For molecules containing only a single quantum proton, the core guess computed within NEO-HF provides the exact solution for the given electron--proton Coulomb interaction $J_{\mathrm{ep}}$. However, since the electronic density is also optimized in NEO-SCF computations and the magnitude of $J_{\mathrm{ep}}$ changes, this solution can still be relatively different from that of the true ground state. Therefore, the use of other initial guesses can be justified in this scenario given that these provide protonic densities closer to those from fully converged NEO-SCF computations. To verify whether this can be the case, $f$-rank scores were computed with the NEO-HF and NEO-DFT methods using Cartesian PB4 basis sets as well as spherical PB4 and PB5 sets employing a stepwise SCF convergence protocol. The obtained data is presented in Section~S3 in the SI. As can be seen from these results, $f$-values produced with a specific combination of the NEO-SCF approach and initial guess do not vary across different basis set types and sizes by more than about 2\%. This can be explained by the fact of the guessed and fully converged protonic densities being weakly dependent on the basis set size and well described even with the smallest PB4-D basis. Furthermore for HOa, HOi and 1s guesses, a single HO-like function is used to represent the guessed density of the quantum proton (before being projected), which is completely independent on the choice of the basis set. This allows us to restrict our consideration to a single example of the Cartesian PB4-F2 basis and analyze it in more details. Results for this case are shown in Figure~\ref{fig:f_ranking}.

\begin{figure}
	\centering
	\includegraphics[width=1.0\textwidth]{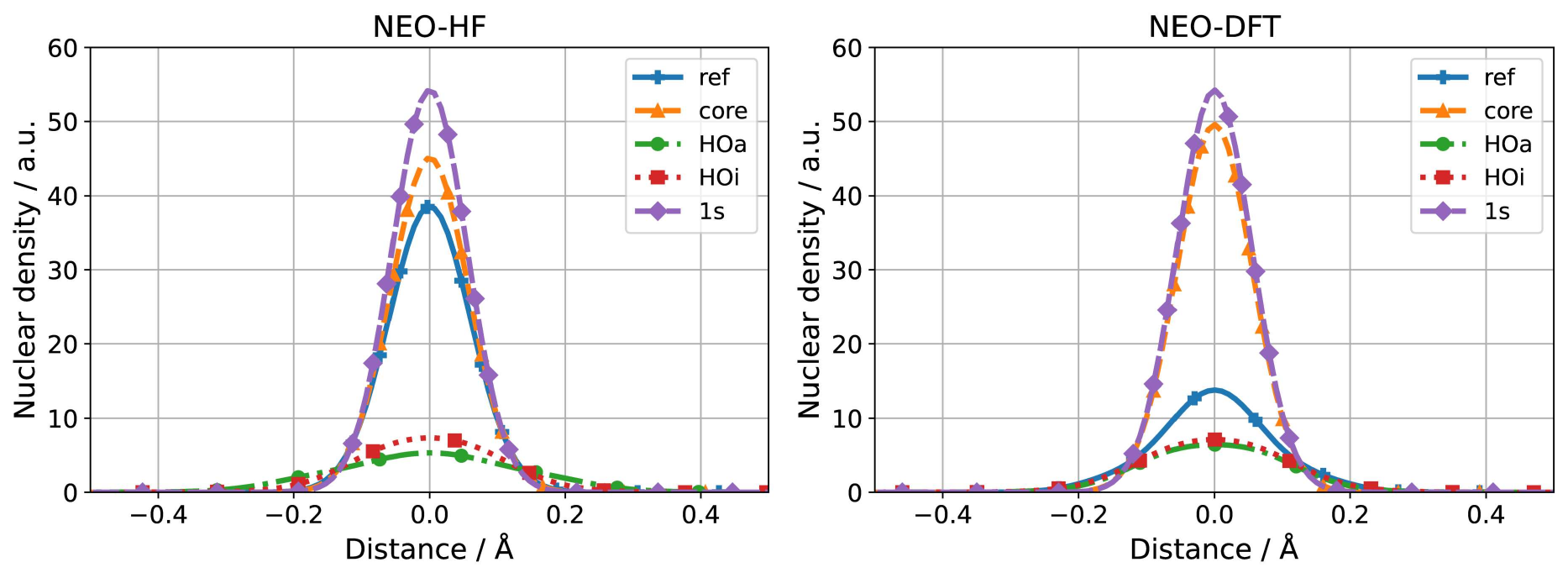}
	\caption{FHF$^-$ protonic density plotted along the F--H--F direction. Results generated with the NEO-HF approach are shown on the left, whereas those from NEO-DFT are on the right. Cartesian PB4-F2 basis set was employed in both cases.}
	\label{fig:fhf}
\end{figure}

As can be seen from Figure~\ref{fig:f_ranking}, performance of initial guesses is very different for the NEO-HF and NEO-DFT methods. In case of NEO-HF, the quality of the guess varies from 0.51 to 0.98 in the sequence HOa $<$ HOi $<$ 1s $<$ core with 1s and core guesses producing considerably better results compared to others. However, for NEO-DFT the best performing guess is HOi having the $f$-value of 0.93, whereas HOa, 1s, and core show a comparable performance with $f$-scores of 0.74--0.77. These results can be attributed to the well-known fact of the NEO-HF method producing overly localized protonic densities~\cite{Sirjoosingh2013,Brorsen2015}. 
It is not surprising that the 1s guess works well for NEO-HF, since it features a large exponent of 16$\sqrt{2} \approx 22.6$~bohr$^{-2}$ and, therefore, produces a very tight s-function. 
For comparison, exponents computed in the HOi approach vary from about 6 for the FHF$^-$ molecule to about 17 for H$_2$O (for the complete list of exponents computed with the HOi approach, see Section~S4 in the SI), therefore resulting in much more delocalized protonic distributions. As for the core guess, in the current implementation it is computed from the given proton--electron Coulomb interaction $J_{\mathrm{ep}}$, which in turn depends on the electronic density of a molecule without quantum protons, possibly resulting in rather localized densities. This assumption is further supported by the analysis of the FHF$^-$ protonic density along the F--H--F direction as presented in Figure~\ref{fig:fhf}. As seen from this example, the fully converged reference density is indeed much more localized in the case of NEO-HF and is better described by the 1s and core initial guesses, whereas the opposite scenario is observed for NEO-DFT with the delocalized reference density and HOa and HOi guesses providing the best description. Very similar results are obtained for molecules featuring multiple quantum protons (see Section~S3 in the SI).

The overall poor/moderate performance of HOa, as seen from Figure~\ref{fig:f_ranking}, might indicates that accounting for the protonic density anisotropy does not necessarily lead to an improved quality of the initial guess, and, as seen from very large range bars, is very molecule dependent. Alternatively, it might be due to an inability of HOa to correctly describe the anisotropy, i.e., small angular frequencies not being reliably computed. Thus, HOa provides very good starting densities in NEO-DFT computations of the FHF$^-$, HNO, HOF molecules with $f$-scores being over 0.9, whereas H$_2$S is poorly described with the score of only 0.3. This poor performance of HOa together with rather high complexity in terms of theory and program implementation make it unpractical. On the contrary, the HOi guess is much easier to implement and shows an excellent performance in NEO-DFT computations. Therefore, it can serve as a good alternative to 1s and core.

Since the exponent $\zeta = 16 \sqrt{2}$~bohr$^{-2}$ used in the 1s initial guess was found to be overly large for NEO-DFT computations, we determined a more optimal value. To that end, NEO-DFT computations for molecules containing a single quantum proton using the PB4-F2 basis set and starting from the 1s guess were performed over a grid of $\zeta$ values ranging from 8 to 22. It was found that the use of $\zeta = 10$ results in an average $f$-rank score of about 0.97, with minimal and maximal values equal to approximately 0.97 and 0.99, respectively.

\subsection{Number of SCF Iterations} \label{sec:num_iters}

It should be noted that using an initial guess, which is closer to the final NEO-SCF solution does not necessarily mean a faster convergence. For example, during the very first nuclear SCF procedure of the stepwise convergence protocol of NEO-HF, the protonic density can be approximated with any initial guess, but then it will converge within subsequent iterations to the best solution for the given proton--electron Coulomb interaction $J_{\mathrm{ep}}$. In the case of a single quantum proton, this solution is equivalent to the density provided by the core guess. In other words, even if different guesses are used in the stepwise procedure, same density is always reached after the very first nuclear SCF run. The number of subsequent iterations could vary slightly purely due to numerical noise and peculiarities of the SCF optimization algorithm. Therefore, all initial guesses, except for core, would result in a comparable number of macro and SCF iterations. In this scenario, only the core guess has a clear advantage over others, since it does not require more than one nuclear SCF iteration per each marco cycle (if implemented and executed accordingly, i.e., without damping or level shifting being enabled).

\begin{figure}
	\centering
	\includegraphics[width=0.9\textwidth]{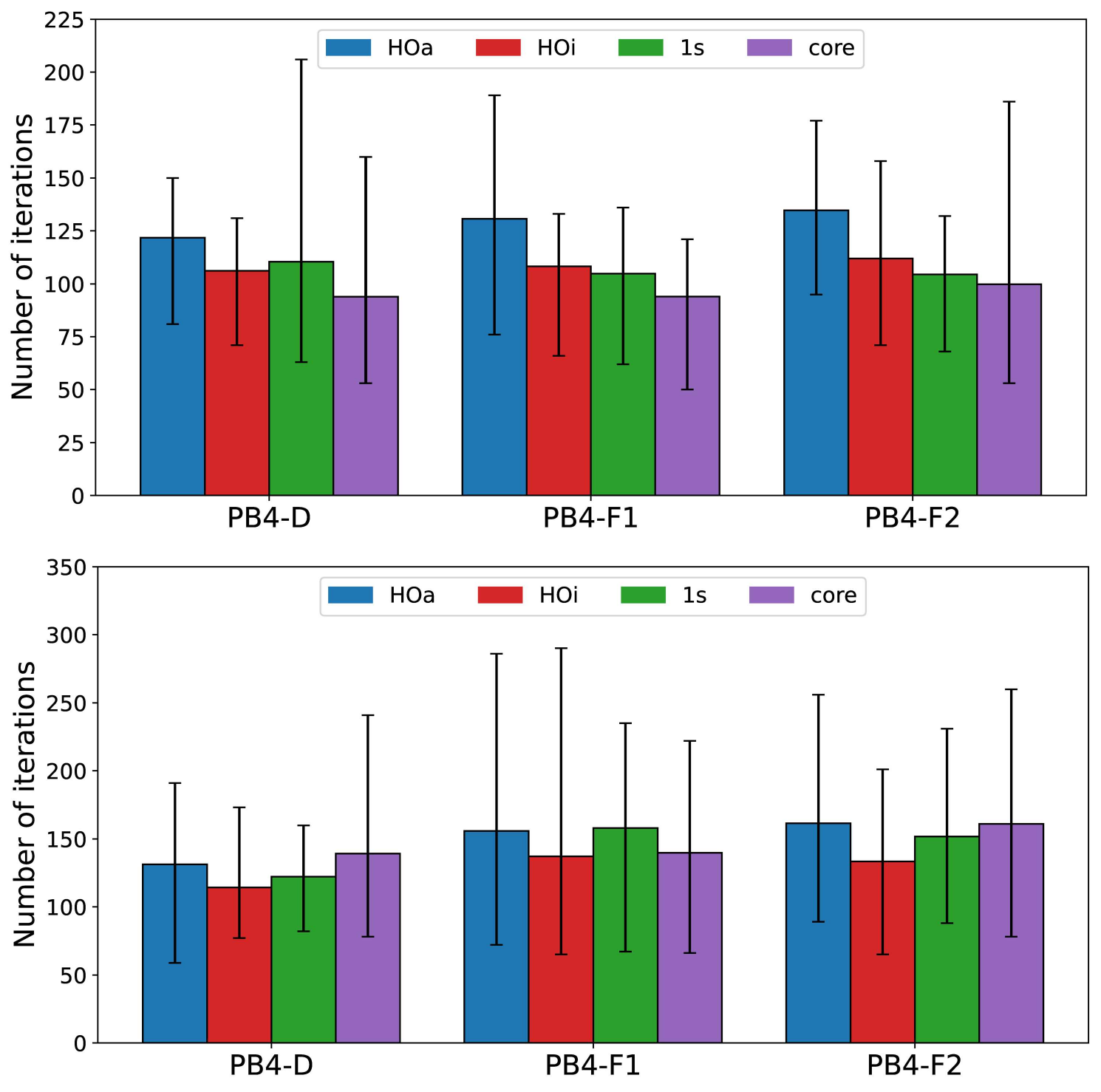}
	\caption{Numbers of SCF iterations required to reach convergence in NEO-HF (top) and NEO-DFT (bottom) computations employing the simultaneous convergence protocol and Cartesian PB4 basis sets. Bars correspond to numbers of SCF iterations averaged over all molecules in the set, whereas range bars indicate minimal and maximal values.}
	\label{fig:scf_iter}
\end{figure}

This line of argumentation does not apply to the simultaneous convergence protocol (as well as to computations of molecules with multiple quantum protons). In this case, the number of SCF iterations was recorded and analyzed. The complete set of obtained results is given in Section~S5 in the SI, whereas an example for Cartesian PB4 bases is shown in Figure~\ref{fig:scf_iter}. Similar to the discussion above on $f$-rank scores, the core guess shows the best performance in all NEO-HF computations producing the smallest number of SCF iterations compared to other guesses, while HOi is superior over other approaches for the large majority of NEO-DFT calculations. The only exceptions from this trend are found in NEO-DFT employing the spherical PB4-F1 and PB4-F2 sets, where the 1s and core guesses show slightly better performance, respectively (see Section~S5 in the SI).

\subsection{Computations of Approximate Hessians} \label{sec:app_hessians}

As was discussed previously, HOi shows superior performance in NEO-DFT computations compared to other initial guesses, while also being relatively easy to implement in a program code. The only disadvantage of this approach comes from the need to compute a partial 3$\times$3 Hessian matrix, which can become computationally very unfavorable for large molecules. However, such Hessian matrices can be evaluated using lower-cost computational approaches. To demonstrate this, a series of computations was performed for H$_2$O$\cdots$\textbf{H}OH and F\textbf{H}F$^-$ molecular systems featuring different degree of density localization (as deduced from the values of exponents obtained with HOi). In these computations, lower-cost approaches were used to optimize molecular structures and generate corresponding partial Hessians. Only the latter was then used in subsequent NEO-DFT computations, while keeping all settings as well as the employed molecular structure unchanged. Results of these computations are demonstrated in Tab.~\ref{tab:app_hessians}.

\begin{table}[h!]
	\small
	\center
	\caption{Exponents $\zeta$ and $f$-scores computed for H$_2$O$\cdots$\textbf{H}OH and F\textbf{H}F$^-$ molecular systems using lower-cost Hessians and the NEO-DFT approach with the spherical PB4-F2 protonic basis set. All computations converged to the same NEO-DFT energy value within an error below about 8.8$\times 10^{-6}$~a.u. Exponents $\zeta$ are in bohr$^{-2}$ units.}
	\label{tab:app_hessians}
	\begin{tabular}{lcccc}
		\hline
		\hline
		Method &  \multicolumn{2}{c}{H$_2$O$\cdots$\textbf{H}OH} &  \multicolumn{2}{c}{F\textbf{H}F$^-$}   \\
		\cmidrule(lr){2-3}
		\cmidrule(lr){4-5}
		       &  $\zeta$ &  $f$-score &  $\zeta$ &  $f$-score  \\
		\hline
		PBE/aug-cc-pVTZ & 14.29 & 0.92 & 5.87  & 0.93    \\
		PBE/cc-pVDZ     & 14.22 & 0.92 & 6.83  & 0.97   \\
		PBE/STO-3G      & 13.86 & 0.93 & 12.63 & 0.94  \\
		GFN2-xTB        & 10.62 & 0.97 & 6.26  & 0.95   \\		  
		\hline
		\hline    
	\end{tabular}           
\end{table}

As one can see, values of exponents $\zeta$ computed for the water dimer H$_2$O$\cdots$\textbf{H}OH using different approximate Hessians do not vary by more than about 4 units (from about 10 to 14). The corresponding $f$-scores show a very weak dependency on $\zeta$ and are larger than 0.90 in all cases consistently providing very good starting guesses for NEO-DFT computations. The case of the F\textbf{H}F$^-$ molecule can be considered as more challenging since accurate computations of anions require diffuse basis functions to be used. Although lowering the basis set size from aug-cc-pVTZ to cc-pVDZ does not strongly affect the value of $\zeta$, the use of the minimal STO-3G basis set results in a more considerable change of about 6 units. However, the corresponding $f$-scores again do not show a strong dependency on $\zeta$ and stay above 0.90 in all cases. Therefore, the use of GFN2-xTB for the generation of approximate partial Hessians can be considered as a route to making the HOi guess inexpensive and practical.

\section{Conclusions and Outlook} \label{sec:conclusions}

In this work, a novel protonic initial guess based on the analytical solution of the three-dimensional HO problem was proposed, its relation to the existing 1s-type guess was established, and its performance was systematically evaluated against existing approaches. 
To this end, both $f$-rank scores, which show the similarity between the guessed and fully converged density matrices, and the numbers of protonic SCF iterations, reflecting the speed of convergence, were analyzed. The anisotropic variant of the HO guess, denoted HOa, was found to be impractical due to its poor performance and the considerable program implementation effort required. In contrast, the isotropic variant, referred to as HOi, outperformed all other initial guesses, when being used in NEO-DFT calculations in conjunction with the simultaneous SCF convergence protocol. Furthermore, it was found to be relatively straightforward to implement. For NEO-HF calculations and stepwise optimization algorithms, however, the one-proton (core) and 1s-type guesses exhibited comparable or slightly better performance than that of HOi. This was attributed to the considerably more localized protonic densities in the NEO-HF method. Subsequently, a reparametrized version of the 1s guess with improved performance in NEO-DFT computations was proposed. Although the majority of discussed results were obtained for molecules containing a single quantum proton, very similar trends were observed for systems featuring multiple quantum protons. The latter were found to be challenging to converge with simple SCF optimization protocols, used in this work.

The main disadvantage of the proposed HOi approximation arises from the need to evaluate partial 3$\times$3 Hessian matrices for each quantum proton, which becomes computationally demanding for large molecular systems. However, it was demonstrated that Hessians computed with lower-cost computational methods such as GFN2-xTB can be employed for this purpose without strongly affecting the quality of the initial guess. As a result, HOi can be applied in a highly efficient and robust manner. However, it should be noted that the presence of imaginary frequencies in, for example, calculations of molecules in transition states, can pose another challenge for the HOi approach. In such cases, HOi could be applied for closely-related minimal molecular structures to obtain an optimal starting density or the 1s function exponent, which can subsequently be re-used in computations of transition states. Alternatively, it can be regarded as a parametrization strategy for the existing 1s-type initial guess, enabling the determination of optimal exponent values for a broad range of molecules, hydrogen bonds, and types of quantum nuclei.

\begin{acknowledgement}
The author thanks Larissa Sophie Eitelhuber for preliminary testing of the NEO-DFT and NEO-HF codes and Iasemin Topchu for the program implementation of the simultaneous DIIS algorithm. Helpful discussions with 
Dr.\ Denis S.\ Tikhonov, Dr.\ Susi Lehtola, 
and Prof.\ Ricardo Mata
on theory and algorithms are greatly appreciated. The author acknowledges funding provided by the German Research Foundation (Deutsche Forschungsgemeinschaft, DFG), project number 545861628, and the HPC Service of FUB-IT, Freie Universit\"at Berlin, for computational time.

\end{acknowledgement}

\begin{suppinfo}

Additional theory aspects on the basis set projection technique and derivations of analytical expressions for overlap integrals are presented in Section~S1. 
In Section~S2, the correctness of the NEO-HF and NEO-DFT program implementations is assessed. 
Raw data supporting the main findings of this work is given in Sections~S3--S5.

\end{suppinfo}

\newpage
\clearpage

\providecommand{\latin}[1]{#1}
\makeatletter
\providecommand{\doi}
  {\begingroup\let\do\@makeother\dospecials
  \catcode`\{=1 \catcode`\}=2 \doi@aux}
\providecommand{\doi@aux}[1]{\endgroup\texttt{#1}}
\makeatother
\providecommand*\mcitethebibliography{\thebibliography}
\csname @ifundefined\endcsname{endmcitethebibliography}
  {\let\endmcitethebibliography\endthebibliography}{}

\end{document}